\documentclass[11pt]{article}
\usepackage{setspace}

\usepackage{longtable}
\usepackage{graphicx}
\usepackage{rotating}
\usepackage{lscape}
\usepackage{multirow}
\usepackage{multicol}
\usepackage{algorithm, algpseudocode, algorithmicx}
\usepackage{wrapfig}
\usepackage{caption}
\usepackage{amsmath}
\usepackage{bbm}
\usepackage{amsthm}
\newtheorem{definition}{Definition}
\newtheorem{theorem}{Theorem}
\newtheorem{corollary}{Corollary}[theorem]
\newtheorem{lemma}{Lemma}
\usepackage{amsfonts}
\usepackage{MnSymbol}
\usepackage{wasysym}
\usepackage{color,soul}
\usepackage{hyperref}

\usepackage{url}
\usepackage{pbox}
\usepackage{graphicx}
\usepackage{wrapfig}
\usepackage{subfigure}
\usepackage[numbers]{natbib}
\usepackage{comment}
\usepackage[title]{appendix}

\usepackage{setspace}

\usepackage{epstopdf}
\AppendGraphicsExtensions{.tif}

\title{
Relaxed Clique Percolation and Disinformation-Resilient Domains for Social Commerce Networks
}
\author{Himangshu Paul, Alexander Nikolaev }
\date{\today}
\begin{document}

\maketitle

\begin{abstract}
\noindent 
Must we trace and block all fake content in a social commerce network so that genuine users may enjoy \emph{fake-free} information? Such efforts largely fail, because, as we get better at spam detection, spammers use the same advances for anti-detection.
As a fundamentally new approach, we show that an online platform can aggregate and route user-generated content in a smart personalized way, which fosters and relies on \emph{``collective social responsibility"}. 
We introduce the notion of \emph{information aggregation domain}, or simply, \emph{domain}:
composed for a given ``central" node (user account), a domain is a connected set of nodes whose user-generated content is eligible to be used to meet the central node's information needs. 
Admitting malicious information sources---``bad citizen" nodes---into ``good citizen" nodes' domains puts the good citizens at risk for disinformation attacks.
We show how a platform can limit this risk by exploiting the social link structure between its nodes without the need to know which nodes are good or bad citizens.
We introduce Relaxed Clique Percolation (RCP), a class of policies to compose personalized disinformation-resilient domains. 
Then, we define ``RCP cores" and show how they can be used to efficiently compose resilient domains for all network nodes at once. Finally, we analyze the properties of RCP domains found in real-world social networks including Slashdot, Facebook, Flickr, and Yelp, to affirm that in practice, RCP domains turn out to be large and spatially diverse. 
\end{abstract}


\section{Introduction}
\footnote{This work is a concise version of the dissertation titled ``Social responsibility and combating disinformation effects in social commerce networks: Composing resilient domains via relaxed clique percolation". Please refer \href{https://www.proquest.com/openview/ae2d50261bc936ee321852b29c10b27a/1?pq-origsite=gscholar&cbl=18750&diss=y}{here} for further details.}
\footnote{Author Contributions: A.N. designed research; H.P. and A.N. performed research; H.P. analyzed data; A.N. and H.P. wrote the paper.}
Online social networks, and more generally, online platforms including Facebook, Twitter, Instagram, Quora, Medhelp, Yelp, Amazon, etc., store and regulate the propagation of posted user-generated content among their users \citep{papanastasiou2018}. An online post may contain a description of an experience, an opinion, a recommendation, 
an answer to a question, a piece of news, or a rumor. The users that get exposed to posted content may disregard it as useless, get a passing enjoyment from it, or use the provided information to support a judgment. 

Online platforms care not to flood their users with information, while keeping them engaged with the peers and content that fits their needs. Most platforms enable users to declare other users' accounts as ``friendly", e.g., define friend-circles on Facebook or trust-circles on Slashdot, create lists of users-to-follow on Twitter, etc. This gives each online social network actor control over the content that serves their social interaction needs. However, the actor's level of control is limited when it comes to information gathering needs. To satisfy such needs (e.g., for product reviews), actors have to rely on the platform to present them with credible information from socially-far-away sources, i.e., from the accounts of actors whom they do not know. Indeed, when delivering user-generated content to a user node, unprompted or in response to a search query, the online social network platform implicitly or explicitly uses a select set of its other nodes as information sources \citep{Almaatouq2020}. 

Let the userbase of an online platform be modeled as a social graph $G=G(N,E)$, where $N$ is the set of nodes (representing users) and $E=\{(i,j): i,j \in N, i\neq j\}$ is the set of links (representing ``friendly" connections). For a clear exposition of the ideas of this paper, we assume that the links are \emph{undirected}, express \emph{mutual trust}-type relationships, and are user-initiated information propagation channels. Consider an information gathering task initiated by node $i$, where the platform has to rely on user-generated content to inform the results of a search/survey/poll, collect (popular) posts with a given tag, generate a news feed, or report an average star-rating of a commercial item. An \emph{information aggregation domain}, or simply, \emph{domain} of this ``central" node $i$, denoted by $D^{(i)} \subseteq N$, is a set of nodes that the platform will use as the (potential) sources of information for $i$. 
The \emph{signals} from these nodes -- be it informative posts, recommendations, or product reviews -- will be collected and presented in a list, or post-processed/aggregated for the viewing convenience of $i$. Note that any particular node in $D^{(i)}$ does not need to be able to contribute to responding to \emph{every} informational query that $i$ submits. Also, a signal may have a fractional probability to reach $i$, as is the case, e.g., with the propagation of Instagram posts into follower feeds.

Domain composition presents a challenge because online platforms of today cannot help but host ``bad citizens" that can manipulate ``good citizens" toward achieving financial or ideological gains. The financial motive is most transparent. Now that social commerce has flourished on LinkedIn, Facebook, Instagram, Twitter, etc., any social network actor can act as a seller offering products/services and as a customer buying and reviewing others' products/services \citep{Liang2011, Stephen2010, Selem2023}. The growth of revenues facilitated by social commerce networks (SCN) motivates immoral vendors to fake and/or buy user-accounts; these bad citizens then post deceptive product reviews and provide highly biased recommendations.
Theoretical and practical efforts in the false information detection and social bot detection areas have been unable to eradicate bad citizens; each online platform continuously works to keep their number below some practically manageable threshold but the volume of the unfiltered ``bad" content remains huge \citep{Paul2021a, papanastasiou2020}.

\begin{figure}[t]
\begin{center}
\includegraphics[height=3in]{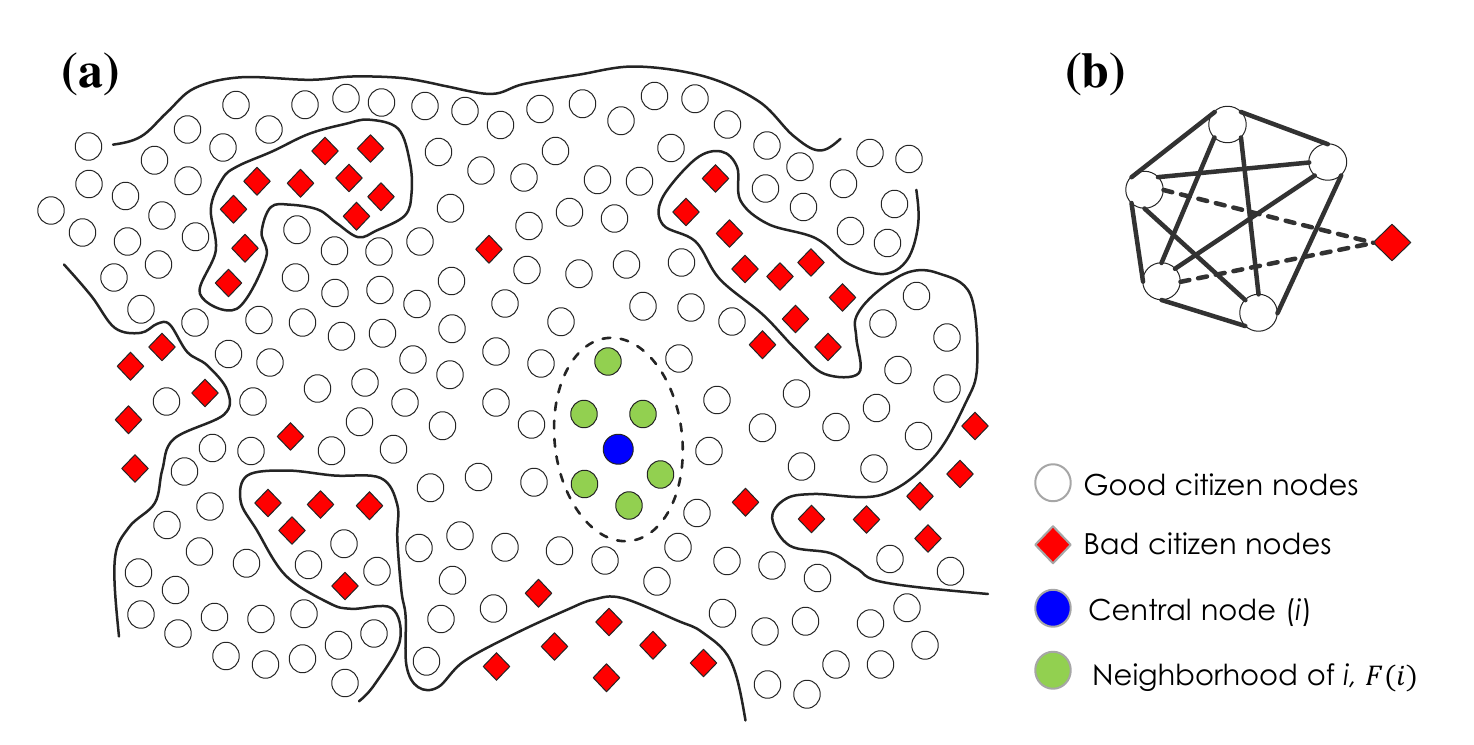}
\caption{(a) Concept of a resilient domain composed for a ``central" node $i$. It is large (extends far), contains mostly good citizens, and bars mass-infiltration of bad citizens. Naturally, the direct neighbors of $i$ are first to get admitted into the domain. (b) Key assumption of socially responsible behavior: while a good citizen can occasionally make a mistake and link with a bad citizen, the case where a well-communicating \emph{group} of good citizens holds the same bad citizen as friend is extremely unlikely.} \label{Figure_1}
\end{center}
\end{figure}

A noble goal for any SCN is to ensure that the domains it composes for its good citizens are resilient against disinformation attacks of spammers. We use the term \emph{disinformation-resilient}, or simply \emph{resilient}, as a property of domain composition policies, and by extension, of domains composed under such policies. A desirable domain composed for a good citizen should (i) contain as many good citizens as possible to ensure availability and diversity of credible information in response to information search queries, and (ii) contain as few bad citizens as possible to minimize disinformation risks. Note that a domain does not have to be completely free of bad citizens to be resilient; a limited fraction of bad citizens in a large domain cannot do much damage. The key is to ensure that a good citizen's domain does not contain such a group of bad citizens \emph{whose voice can supersede the voice of the domain's good citizens} when the platform aggregates search query results over the entire domain.
To sum up, practically useful resilient domains should be large, i.e., extend far beyond one's friend-circle, while bypassing any bad citizen conglomerates in the social graph, as schematically shown in Figure \ref{Figure_1}(a).

This paper assumes no available ``ground truth" information that could ascertain a SCN node as good citizen. We take it that confirmed bad citizens are at once eliminated from a SCN by the spam detection algorithms; all the other nodes can be assumed ``likely good" but with no guarantee. We henceforth proceed by making assumptions on link formation behavior of socially responsible humans. 

We assume first that a friend-circle of a good citizen $i$, $F(i)= \{j: (i,j)\in E\}$, can be expected to mostly contain good citizens, which are credible information sources at least in the sense that they are accounts of real people. Hence, assuming that a given central node is a good citizen, one can use its nominations of ``friendly" peers as the grounds for composing a resilient domain. Indeed, if a good citizen were to obtain an aggregate response to a query, they would certainly agree to poll all their friends for answers and ``average out" those answers.
Note, however, that a domain composed by simple extension of friends-of-friends' circles -- in the form $D^{(h)} = \bigcup \{ J = \{ j: j\in F(i)\} \}\bigcup\{ K = \{ k: k\in F(j), j\in J \} \}\bigcup ...$ -- is unlikely to stay resilient. 
This is because links between good citizens and bad citizens do occur, however rarely; hence, a domain in the above form may eventually extend into a bad citizen's friend-circle, and thereafter, an arbitrarily large number of bad citizens can be admitted into the domain. Indeed, adding new bad citizens to a SCN and linking them to already-owned bad citizen accounts comes at little cost to an opinion spam farm. More generally, the just-described ``mass-infiltration" of bad citizens into a domain is a result of ``breach" events when the domain composition policy admits a bad citizen that is not linked with any of the domain's good citizens, i.e., when bad citizens by themselves can control the subsequent admittance of nodes into the domain. 

We assume next that bad citizens cannot ``stay undiscovered" in the midst of densely-connected good citizen friendship groups. This is because, while a single good citizen may accidentally admit a bad citizen as friend, the chance that multiple linked (i.e., effectively communicating) good citizens all make the same mistake is negligible. Clique, a subgraph where all nodes are directly linked to each other, enforces complete connectivity of members of a social group. Generalizing this strict connectivity requirement, we posit that if a cohesive subgraph -- a defined \emph{relaxed clique} -- is known to contain a certain large fraction of good citizens, then all of its nodes must be good citizens. Then, one way to compose a resilient domain for a good citizen central node is to take its friend-circle and \emph{sequentially expand it by overlapping cohesive subgraphs} with pre-imposed cohesion properties. We argue that domains composed in this way can be both resilient \emph{and}, with high probability, large.

The objective of this paper is to present a theoretical basis and algorithmic approach toward composing \emph{resilient domains}. To this end, we rely on the \emph{clique relaxation} and \emph{clique percolation} concepts. Clique relaxation is a mathematically sound approach for defining small cohesive subgraphs. The Clique relaxation idea makes cliques even more useful for modeling human social groups by relaxing three elementary cohesiveness properties of a clique: familiarity, reachability, and robustness \citep{Pattillo2013a}. For instance, distance-based relaxed clique models, e.g., $s$-clique and $s$-club, relax the \emph{reachability} property so that the members of such subgraphs need not be within one-hop away from each other: instead, they can be maximum $s$-hops away ($s\in \mathcal{I}^+$). {Clique percolation}, due to Palla et al., is a modeling idea where overlapping small cohesive subgraphs ``cover a network" in a diffusion-like manner to iteratively and systematically ``grow" and ``separate" social communities \citep{Palla2005, Derenyi2005}; it falls under the umbrella of community detection research \citep{Newman2002, Newman2003, Fortunato2010}. The above ideas come together in our Relaxed Clique Percolation (RCP) logic, which supports a dynamic -- community-building -- view on resilient domain composition. 

First, within the flow of the paper, we present RCP as a domain expansion policy. Toward composing resilient domains, the RCP policy prescribes finding connected subgraphs whose cohesion-based properties prevent mass-infiltration of bad citizens in domain expansion. This expansion is sequential. Think that a domain is being constructed iteratively: given some nodes already in the domain, we find a subgraph (with specific cohesiveness property) that overlaps with the domain (has some fraction of nodes in common with the domain's nodes); if the overlap is sufficient, then all the nodes of the subgraph become part of the domain. Thus, the domain expands by percolation of the subgraphs. We refer to the fundamental concepts from the clique relaxation theory for defining and arguing about the cohesiveness properties of the useful subgraphs. This logic can be applied to identify a largest guaranteed-resilient domain for any given central node.

Second, we study RCP policies in application to composing resilient domains under realistic assumptions of a particular parametric form. We prove, under such assumptions, that a domain built using a tuned RCP policy, is resilient against mass-infiltration of bad citizens and disinformation attacks.

Third, we introduce the notion of ``RCP core'' and, from a macroscopic network outlook, show how resilient domains can be efficiently composed for all the nodes of a large social network in one pass. 

Fourth, we use RCP policies to compose domains for nodes in a number of samples from real-world networks.
We investigate the influence of two factors on the sizes of built resilient domains: (i) assumptions on good citizen behavior, and (ii) availability of link information.

The value of studying domains, i.e., specifying how they should be composed, lies in providing a calculated, personalized approach to information retrieval, which prevents disinformation spread to socially responsible SCN users while providing them with access to a sufficient number of information sources to execute queries. This approach combats the \emph{impact} of disinformation without having to directly detect it. 

\section{Domain Composition and Resilience}
In this section, we first state the assumptions on socially responsible good citizen behavior. Next, we define domain, trivial domain and domain resiliency, and then, domain expansion. We introduce expansion feasibility properties that impose restrictions on domain expansion, and finally, introduce RCP as a kind of sequential domain expansion that is governed by the logic of clique relaxation and clique percolation theories.

\subsection{Assumptions on Good Citizen Behavior}

This section presents the clauses that describe behavior of good citizens, stating them as \emph{assumptions}. One can interpret such clauses as \emph{expectations} or \emph{standards} of socially responsible behavior. These standards may not be maintained by all humans in real-world online social circles \emph{today} because online platform users are not necessarily aware of indirect ill-effects of their irresponsible link-building behavior. Meanwhile, once ideas and algorithms such as those proposed in this paper get applied in practice, then people will value the integrity of their social circles more, as those circles will tangibly impact the results of their informational queries. In this light, we believe it will be easy for the reader to accept the stated assumptions. Note that they are all parametric, which adds flexibility and realism. 

Consider a social graph $G=G(N,E)$, where $N$ is the set of nodes and $E=\{(i,j): i,j \in N, i\neq j\}$ is the set of links. The set $N$ has two mutually exclusive and collectively exhaustive subsets: subset $H$ of good citizens (``Humans") and subset $B$ of bad citizens (``Bots"). We view $G$ as an outcome -- more precisely, a snapshot in time -- of the process of emergence of social actor nodes (online SCN accounts) and links, in which both good citizens and bad citizens participate. However, to the onlooker, it is unknown which particular nodes are good or bad citizens. In other words, for this onlooker, as well as for any node viewed as a SCN user account holder, there is uncertainty about the genuineness of the (other) SCN  nodes, i.e., uncertainty about their belonging to the sets $H$ or $B$.  

First, we look at \emph{individual} behavior of good citizens and friend-links they tend to build. It is natural to assume that a socially responsible human will link to the accounts whose owners they know personally \citep{borgatti2003}. To express the likelihood of accidental misjudgments, we introduce the concept of ``(social) responsibility threshold" $0<r<<1$, inversely proportional to the level of care that a socially responsible human maintains while building SCN friendship links.

Next, we look at how good citizens cluster together. Here, one assumption describes the behavior of good citizens as they form \emph{small groups}: in particular, we expect that triadic closures should occur less often in ``two good citizens, one bad citizen" triads. In other words, while it is already unlikely for a good citizen to have a bad citizen in their neighborhood, it is even more unlikely that the good citizen will have common friends with that bad citizen. Another assumption that concerns \emph{collective} behavior is about membership within \emph{yet-larger groups} of good citizens. Connected good citizens are assumed to communicate with each other and share information about their social contacts: once a good citizen raises a suspicion about a certain account, then their connected good citizen peers follow suit. Thus, a large connected group of good citizens will never keep a bad citizen as all-common friend. 

The assumptions are formally stated as follows:

\begin{itemize}
    \item Assumption 1 ($A1$): 
    Given that node $i\in N$ is a good citizen linked with node $j\in N$, the probability that $j$ is a bad citizen is bounded by the responsibility threshold (r): $P(j\in B |  i\in H, (i,j)\in E) \leq r$. This probabilistic statement is formulated as an inequality to reflect the fact that the probability of having a bad citizen as friend is at the highest ($r$) when $i$ has only private information about $j$; this probability may be lower than $r$ when $i$ has additional, network peers-dependent information about $j$; but if such additional information raises suspicion of $i$ toward $j$ to increase the probability beyond $r$, then $i$ is assumed to immediately sever the link $(i,j)$, and hence, such links are not found in the social graph.
    \item Assumption 2 ($A2$): Given that node $i\in N$ is a good citizen that is linked with node $j\in N$ and has at least $x\in \mathcal{I}^+$ mutual friends with $j$, then node $j$ must be a good citizen: $P(j\in B  i\in H, (i,j)\in E, |F(i) \cap F(j)|>=x) = 0$. Note that under this assumption, any clique of size equal to or larger than $x+2$ found in the social graph must consist either entirely of good citizens or entirely of bad citizens.
    \item Assumption 3 ($A3$): Let $H^{\leftrightarrow} \subset H$ denote a \emph{connected} set of good citizens. Given that node $j\notin H^{\leftrightarrow}$ is linked with every member of $H^{\leftrightarrow}$, and the size of $H^{\leftrightarrow}$ is at least $y\in \mathcal{I}^+$, then node $j$ must be a good citizen:
    $P(j\in B | F(j) \supseteq H^{\leftrightarrow}, |H^{\leftrightarrow}|\geq y ) = 0$. Note that under this assumption, any clique of size equal to or larger than $y+1$ found in the social graph must consist either entirely of good citizens or entirely of bad citizens.
\end{itemize}

\noindent To recap, $A1$ postulates the ability of a human to self-defend against disinformation;
$A2$ expresses how the individual defensive ability strengthens in small groups of humans, as Grannovetter's ``strong ties" get formed \citep{Granovetter1973} (see examples in Figure \ref{Figure_3});
$A3$ postulates the strength of defense in large connected groups of socially responsible humans. 

Note finally that in order for $A2$ and $A3$ to be congruent with each other, one must have it that $y\geq x+1$.

\subsection{Definitions: Domains and Domain Resiliency}

Consider node $i \in N$; recall that $F(i)= \{j: (i,j)\in E\}$ is the friend-circle of $i$.

\begin{definition}\label{Ch3_def1}
    (Domain and Trivial Domain) Defined for a given central node $i \in N$, a domain $D^{(i)}\subset N$ is a connected set of nodes that contains $i$. The practically useful \emph{trivial} domain $D^{(i)}_t \equiv i \cup F(i)$ is the ``friend-circle-restricted" domain for $i$.
\end{definition}

A node in $D^{(i)}$ is called a ``member" of $D^{(i)}$; a node outside $D^{(i)}$ is called a ``non-member". The trivial domain for any central node is unique. 
Any larger domain can be specified as a list of the nodes, or node sets, that form the domain. 

\begin{definition}\label{Ch3_def2}
    (Domain Resiliency) Domain $D^{(i)}$ is resilient iff, conditional on $i\in H$, the expected fraction of bad citizens in it is below the responsibility threshold, i.e., 
    \begin{equation}
    \frac{1}{|D^{(i)}|} \cdot \mathbb{E}(\sum_{j\in D^{(i)}} \mathbf{1_B}(j) \mid i\in H) < r,
    \end{equation}
    where $\mathbf{1_B}(j) = 1$ if $j\in B$ and 0 otherwise; $0 \leq  r<< 1-r \leq 1$.
\end{definition}

Clearly, this condition holds for the friend-circle-restricted domain, i.e., for trivial domain, $D_t^{(i)}$, of any good citizen $i\in H$. Indeed, by definition of $\mathbf{1_B}(j)$, for any conditioning variable $K$, one has $\mathbb{E}(\mathbf{1_B}(j)| K) = P(j\in B| K)$, and hence,

\begin{equation}
\label{eqn_trivial_resilient}
\begin{split}
    \frac{1}{|D_t^{(i)}|}\cdot \mathbb{E}(\sum_{j\in D_t^{(i)}} \mathbf{1_B}(j) \mid i\in H) & = \frac{1}{|D_t^{(i)}|} \Big( \mathbb{E}\Big(\mathbf{1_B}(i) \mid i\in H\Big) + \sum_{j\in F(i)} P(j\in B \mid i\in H, (i,j)\in E )\Big) \\
    & = \frac{0+r(|D_t^{(i)}|-1)}{|D_t^{(i)}|} < r.
\end{split}
\end{equation}

\noindent Definition \ref{Ch3_def2} states that a domain composed for a good citizen  node is resilient as long as it provides this node with the same level of assurance of being protected against disinformation as good citizens have about own friend-circles.
Note that, while one can employ the \emph{same} rules to compose domains for \emph{all} nodes in a graph, the definition of resilience is formulated only for domains composed for good citizens. We do not assume or intend to uncover any knowledge about which nodes are good or bad citizens; instead, all our definitions, assumptions, and results are conditional. Our goal is to find such a way to compose domains that the resultant domains are resilient for the \emph{good} citizens, wherever they find themselves in the graph.

\subsection{Domain Expansion and Relaxed Clique Percolation}
The trivial domain of a node, while being disinformation-resilient, will contain too few information sources to satisfy the node's needs in querying the SCN. With the knowledge of the link structure in the SCN graph, larger domains can be composed through ``domain expansion", as some non-member friends of the domain member(s) are added to the domain: these non-members can be viewed as \emph{candidates} for admission into the domain, and these members can be viewed as \emph{sentinels} empowered to protect the domain from bad citizen infiltration. 

\begin{definition}\label{Ch3_def3}
    (Domain Expansion) Consider a domain, $D^{(i)}_1$, a sentinel set $R \subseteq D^{(i)}_1$ and a \emph{connected} candidate set $Q \nsubseteq D^{(i)}_1$, $Q \neq \emptyset$, such that $\forall x \in Q$, $\exists y \in R: (x,y) \in E$. Let node set $\mathbb{X}^{\textbf{P}}_{(R,Q)}(D^{(i)}_1)$ be called expansion of domain $D^{(i)}_1$, where $``\mathbb{X}"$ is the \emph{expansion operator}, the duplet $(R,Q)$ is \emph{(expansion) specification duplet}, with $R\cup Q$ called \emph{expansion set}, and $\textbf{P}$ are expansion feasibility properties formulated as conditions on the elements of the specification duplet. If the specification duplet satisfies the expansion feasibility properties, then we say that domain $D^{(i)}_1$ expands into domain $D^{(i)}_2 = \mathbb{X}^{\textbf{P}}_{(R,Q)}(D^{(i)}_1) = D^{(i)}_1 \cup Q$, and call $D^{(i)}_2$ an expansion of $D^{(i)}_1$. If the specification duplet does not satisfy the expansion feasibility properties, then $\mathbb{X}^{\textbf{P}}_{(R,Q)}(D^{(i)}_1) = \emptyset$. For notational convenience, $\mathbb{X}^{\textbf{P}}_{(\cdot,\cdot)}(\emptyset)\equiv \emptyset$.
\end{definition}

One can compose a domain for a given central node via sequential expansion. Given a static social graph, the method of constructing a sequential expansion is iterative. 

\begin{definition}\label{Ch3_def4}
    (Sequential Expansion of Domain) Consider a domain, $D^{(i)}_1$, and a set of expansion feasibility properties $\textbf{P}$. Let node set $\mathbb{X}^{\circlearrowright \textbf{P}}(D^{(i)}_1)$ be called sequential expansion of domain $D^{(i)}_1$, obtained by a telescopic sequence of expansions, all with the same expansion feasibility properties $\textbf{P}$: $\mathbb{X}^{\circlearrowright \textbf{P}}(D^{(i)}_1) = \mathbb{X}^{\textbf{P}}_{(R_k,Q_k)}(\mathbb{X}^{\textbf{P}}_{(R_{k-1},Q_{k-1})}(\mathbb{X}^{\textbf{P}} \, ... \,\,\, \mathbb{X}^{\textbf{P}}_{(R_1,Q_1)}(D^{(i)}_t)))$.
\end{definition}

An unrestricted sequential expansion  (with $\textbf{P} \equiv \emptyset$) will return the largest connected set of nodes reachable from the central node, without any guarantee that the resultant domain is resilient. Thankfully, one can tailor the expansion feasibility properties toward composing large \emph{resilient} domains. 

We use the concepts of clique relaxation theory to define expansion feasibility properties. We require that each expansion must maintain high cohesiveness: the expansion set ($R\cup Q$) must be a relaxed clique that possesses both high-\emph{reachability} and high-\emph{familiarity} properties (these terms are introduced and used in \citep{Pattillo2013a}); the sentinel set $R$ must be well-connected with (rooted in) the rest of the domain; and the candidate set $Q$ must be well-connected with the sentinel set $R$. 

Viewed as a process, a sequential expansion of a domain can also be called subgraph percolation: the domain grows by addition of overlapping subgraphs. Indeed, we can call the resultant sequential expansion ``relaxed clique percolation" as long as the expansion feasibility properties are informed by the clique relaxation theory. Next, we will see that RCP can be used to produce disinformation resilient domains.

\section{RCP Policy for Resilient Domains}
In this section, we present RCP policy, a sequential domain expansion policy for composing domains. We prove that a domain, built for a good citizen using the parametric RCP policy, is indeed resilient against bad citizen mass-infiltration, and hence, disinformation attacks, under the stated assumptions on responsible good citizen behavior. 

\underline{\textbf{RCP policy}} RCP policy $\pi\equiv \pi(\alpha,\beta)$, with $\alpha, \beta \in \mathcal{I}^+$, is a method, consisting of two \emph{Steps}, of composing a domain for a given central node:
\begin{enumerate}
    \item[] Step 1: Compose a \emph{RCP backbone} $\Bar{D}^{(i)}$ defined as a non-empty sequential expansion $\mathbb{X}^{\circlearrowright \textbf{P}^{\pi}}(\{i\})$ with relaxed-clique-based feasibility properties $\textbf{P}^{\pi}=\{P^{\pi}_1, P^{\pi}_2, P^{\pi}_3\}$:
    \begin{enumerate}
    \item[] $(P^{\pi}_1)$ The expansion set ($R\cup Q$) is a connected set and fully contained within the friend circle of one of its nodes -- \emph{key node} $l$ -- defined that $l \in R\cup Q$ and $(R\cup Q)\setminus \{l\} \subseteq F(l)$ (Note: the key node $l$ can either be in the sentinel set, or in the candidate set.); AND
    \item[] $(P^{\pi}_2)$ The sentinel set $R$ is a connected set that:
        \begin{itemize}
            \item has the key node $l \in R$ where $1 \leq |R| < \alpha$, OR
            \item $|R| \geq \alpha$ where $l \in Q$; AND
        \end{itemize}
    \item[] $(P^{\pi}_3)$ The candidate set $Q$ is well-connected with the sentinel set $R$ such that: 
    \begin{itemize}
        \item for all $q \in Q$, $|F(l) \cap F(q)| \geq \beta$ where $l \in R$, OR
        \item every node $\forall q \in Q$ is a key node. 
    \end{itemize}
\end{enumerate}
    \item[] Step 2: Expand the backbone $\Bar{D}^{(i)}$ obtained in Step 1 with the friend circles of all members of the backbone (not yet in the domain) to obtain a \emph{complete domain}: $$D^{(i)} = \bigcup_{m\in \Bar{D}^{(i)}} \big( m \cup F(m) \big).$$
\end{enumerate}

Note that the domain composed in Step 2 also is a sequential expansion, where the candidate sets are parts of the friendship circles of backbone nodes; hence it is correct to call the entire RCP policy \emph{expansion policy}. 
The next definition introduces ``policy compliance" as the term to describe a relationship between a given domain and a given RCP policy: a non-empty domain that can be composed following Steps 1-2 under the specifications of RCP policy $\pi$ is called $\pi$-compliant domain.

\begin{definition}\label{Ch3_def5}
    (RCP-compliant backbone and RCP-compliant complete domain)
A domain $\Bar{D}^{(i)}$ is $\pi$-compliant backbone if there exists a sequential expansion such that $\mathbb{X}^{\circlearrowright \textbf{P}^{\pi}}(\{i\}) = \Bar{D}^{(i)}$, with some ordered set of specification duplets, $(\{i\},Q_1), (R_2,Q_2), (R_3,Q_3), ...$.
A domain $D^{(i)}$ is $\pi$-compliant domain if there exists a $\pi$-compliant backbone $\Bar{D}^{(i)}$ such that $D^{(i)} = \bigcup_{m\in \Bar{D}^{(i)}} \big( m \cup F(m) \big)$.
\end{definition}

\begin{figure}[t]
\begin{center}
\includegraphics[height=3in]{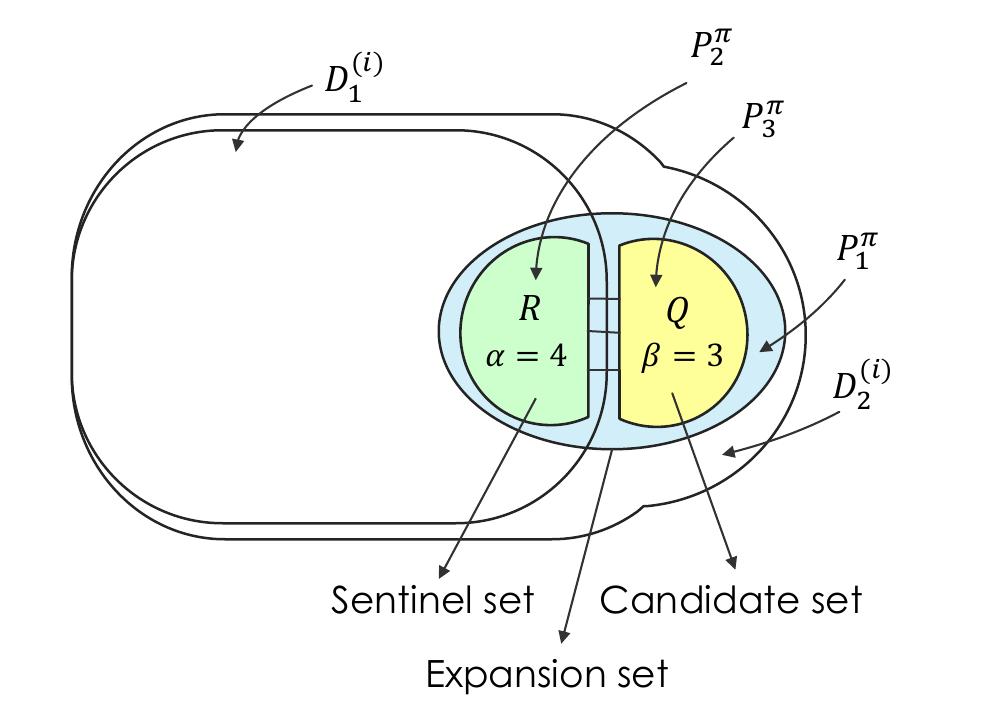}
\caption{Schematic illustration of \emph{one expansion} of a backbone in Step 1 of RCP policy $\pi$. (a) The expansion feasibility properties $\textbf{P}^{\pi}=\{P^{\pi}_1, P^{\pi}_2, P^{\pi}_3\}$ impose necessary conditions on the specification duplet $(R,Q)$: $\{P^{\pi}_1 \}$ restricts $R\cup Q$ to be a relaxed clique that possesses both \emph{reachability} and \emph{familiarity} properties; $\{P^{\pi}_2 \}$ restricts set $R$ to be well-rooted in the rest of the domain; and $\{ P^{\pi}_3 \}$ restricts set $Q$ to be well-connected with set $R$. For example, in order to satisfy $P^{\pi(\alpha,\beta)}$ with $\alpha=4$ and $\beta=3$, set $R$ must have at least 4 connected nodes that have a common friend with set $Q$, or each node in $Q$ must have at least 3 mutual friends with a node in $R$.}
\label{Figure_2}
\end{center}
\end{figure}

\begin{figure}[t]
\begin{center}
\includegraphics[height=2in]{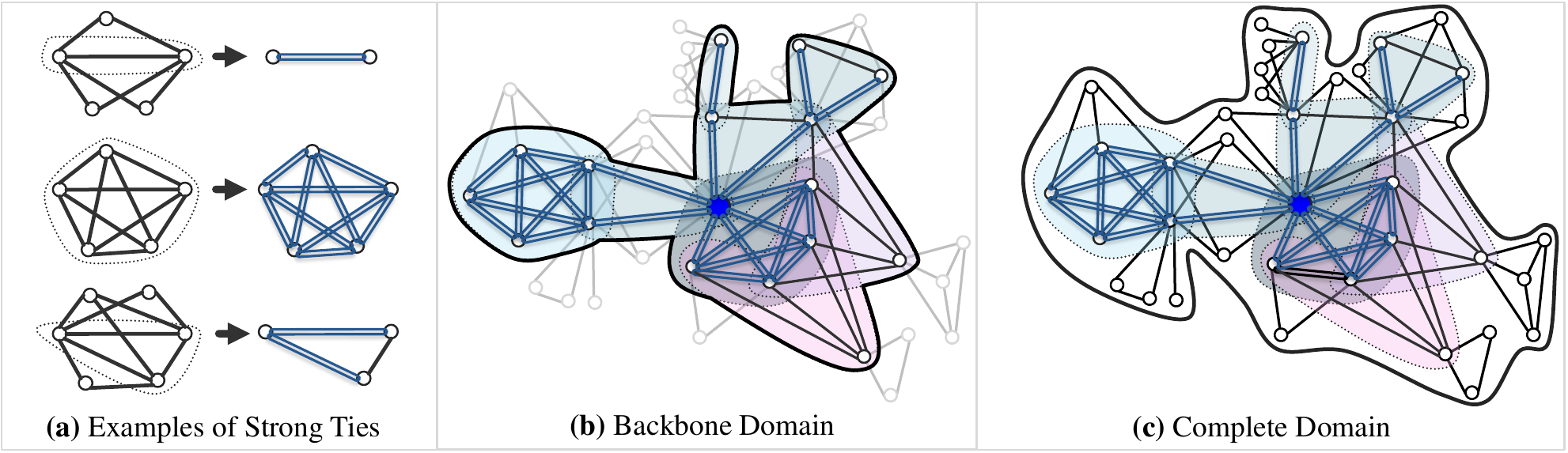}
\caption{
(a) Examples of strong ties. In the picture, the links of strength 3 are depicted as double blue lines; (b) Schematic illustration of a sequential process of composing a backbone of a $\pi$-compliant-domain under some the RCP policy $\pi(\alpha,\beta)$ with parameters $\alpha=4$ and $\beta=3$. The regions bounded by dotted lines are feasible expansions as per $\textbf{P}^{\pi}=\{P^{\pi}_1, P^{\pi}_2, P^{\pi}_3\}$; (c) The complete domain with the backbone in (b).}\label{Figure_3}
\end{center}
\end{figure}

Figure \ref{Figure_2} illustrates a single expansion of a domain backbone under an RCP policy. Figure \ref{Figure_3} illustrates how a $\pi$-compliant-domain is composed, for the RCP policy $\pi(4,3)$.

Next, we prove that the RCP policy $\pi(\alpha,\beta)$ is a resilient domain expansion policy, i.e., that it composes resilient domains for good citizens, under the assumptions of responsible good citizen behavior $A$1-3 with parameters $x$ and $y$.

\subsection{Domain Resilience for Good Citizens under RCP Policy}
With the proofs found in the Supplemental Materials section, first, Lemma \ref{Ch3_lemma3} states that a $\pi$-compliant backbone domain, $\Bar{D}^{(h)}$, composed for a good citizen central node, $h \in H$, contains only good citizens. Then, Theorem \ref{Ch3_theorem1} states that a $\pi$-compliant-domain of $h \in H$ is indeed a disinformation-resilient domain.

\begin{lemma}\label{Ch3_lemma3}
    Assume $A$1-3 hold for social graph $G$. Consider an RCP policy $\pi \equiv \pi(\alpha,\beta)$ with $\alpha \geq y$ and $\beta \geq x$. A $\pi$-compliant backbone composed for a good citizen $h \in H$ does not contain any bad citizens.
\end{lemma} 

\begin{theorem}\label{Ch3_theorem1}
    Assume $A$1-3 hold for social graph $G$ and consider RCP policy $\pi \equiv \pi(\alpha,\beta)$ with $\alpha \geq y$ and $\beta \geq x$. A $\pi$-compliant complete domain $D^{(h)}$ composed for $h \in H$ is disinformation resilient.
\end{theorem}

Note that larger resilient domains serve information needs of any good citizen better. Therefore, it is desirable to compose the largest RCP-compliant domains for all nodes and to do this efficiently, taking the macroscopic view of a social graph as opposed to composing domains for its nodes one at a time. To help this cause, we introduce the concept of RCP core.

\section{RCP Cores and Efficient Domain Composition}
A resilient domain, composed under an RCP policy, is a union of a backbone and the friendship circles of the members of the backbone (by Definition \ref{Ch3_def5}). In a human social network, friend circles of adjacent nodes typically intersect, i.e., friends tend to have many common friends. Consequently, RCP domains of mutual friends largely overlap. This is true even for \emph{non-adjacent} nodes connected by short network paths. Therefore, it may be possible to efficiently compose domains for multiple central nodes at once. To explore this possibility, we first introduce a new type of relationship between node pairs, that of ``a node belonging in another node's domain". This is a \emph{directed} relationship; of special interest is the symmetry of this relationship, which gives rise to the concept of ``RCP core". 

An RCP core is such a node subset where all the nodes find themselves in each others' RCP-compliant backbones. We will show how it comes useful for efficient composition of domains for all nodes in a given social network.

\begin{definition}\label{Ch3_def6}
    (RCP-Core and Supercore)
    Given a set of $\pi$-compliant backbones composed for some or all nodes in $N$ in Step 1 of RCP policy $\pi$, a node set $C^{\pi}$ is a (backbone) $\pi$-core if $\forall c \in C$, $C \subseteq \bar{D}^{(c)}$. A $\pi$-supercore $C^{\pi^*}$ is a $\pi$-core that is not a subset of any larger $\pi$-core.
\end{definition}

Clearly, each node $i \in N$ in graph $G$ is part of at least one resilient domain and is part of at least one $\pi$-core: the set which includes only the node itself is both its domain and a $\pi$-core. It turns out that the entire node set $N$ of graph $G$ can be uniquely split into $\pi$-supercores. 

\begin{theorem}\label{Ch3_theorem2}
    (Existence and Uniqueness of RCP-Supercore Partition)
    There exists a unique partition of node set $N$ of graph $G$ into a set of mutually exclusive and collectively exhaustive $\pi$-supercores $\textbf{C}^{\pi^*}\equiv \{C^{\pi^*}_i \}_{l=1,...,L}\,$, with $L\leq N$.
\end{theorem}

Next, we introduce RCP-supercore-digraph.

\begin{definition}\label{Ch3_def7}
    (RCP-supercore-digraph) Given RCP policy $\pi(\alpha, \beta)$ and graph $G = G(N,E)$, the $\pi$-supercore-digraph $G^{\pi}$ is a directed graph with the node set $\textbf{C}^{\pi^*}$ (i.e., the set of all $\pi$-supercores) and edge set $E^{\pi}$ such that for any pair of $\pi$-supercores, $C^{\pi^*}_1, C^{\pi^*}_2\in \textbf{C}^{\pi^*}$, a directed edge $(C^{\pi^*}_1, C^{\pi^*}_2)\in E^{\pi}$ iff $C^{\pi^*}_1$ contains at least one such set of connected nodes $r_1, r_2, ..., r_\alpha \in N$ and $C^{\pi^*}_2$ contains at least one such node $m \in N$ that $(r_1,m), (r_2,m), ...,(r_\alpha,m)\in E$.
\end{definition}

RCP-supercore-digraphs come useful in constructing the largest RCP-compliant backbones for nodes in $G$.
\begin{lemma}\label{Ch3_lemma4}
     Given RCP policy $\pi(\alpha, \beta)$ and graph $G = G(N,E)$, consider the $\pi$-supercore-digraph $G^{\pi}=G^{\pi}(\textbf{C}^{\pi^*}, E^{\pi})$ and its two connected nodes, $\pi$-supercores $C^{\pi^*}_1, C^{\pi^*}_2\in \textbf{C}^{\pi^*}$ such that  $(C^{\pi^*}_1, C^{\pi^*}_2)\in E^{\pi}$. Then, the largest $\pi$-compliant backbone composed for any node $i\in N$, $i\in C^{\pi^*}_1$, contains all the nodes in $C^{\pi^*}_1$ and all the nodes in $C^{\pi^*}_2$.
\end{lemma}

\begin{theorem}\label{Ch3_theorem3}
     RCP-supercore-digraph is acyclic. 
\end{theorem}
 
\begin{corollary}\label{Ch3_corollary1}
      Given RCP policy $\pi(\alpha, \beta)$, graph $G = G(N,E)$, and $\pi$-supercore-digraph $G^{\pi}=G^{\pi}(\textbf{C}^{\pi^*}, E^{\pi})$, the largest $\pi$-compliant complete domain for any node $i\in N$ such that $i\in C^{\pi^*}\in \textbf{C}^{\pi^*}$ can be composed as follows: (1) obtain the largest $\pi$-compliant backbone for $i$ as the union of $C^{\pi^*}$ and all $\pi$-supercores reachable from $C^{\pi^*}$ via a directed path in $G^{\pi}$, and then, (2) obtain the $\pi$-compliant complete domain for $i$ by applying Step 2 of RCP policy $\pi$ to the obtained backbone.
\end{corollary}

In practice, given RCP policy $\pi(\alpha, \beta)$, we recommend the following method for identifying RCP-supercores in a social graph and constructing its RCP-supercore-digraph:

\begin{enumerate}
    \item[] Step 1: Take graph $G(N, E)$. Build graph $G'$ with nodeset $N'=N$ and edgeset $E'$ defined as follows: for $\forall i',j'\in N'$, an edge $(i',j')\in E'$ exists \emph{iff} $strength(i,j) \geq \beta$. Now find the set of all connected components of $G'$ and denote them as $H'_1, H'_2, ..., H'_K$.
    \item[] Step 2: Build digraph $G''$ with nodes $n_1, n_2, ..., n_K$ and edgeset $E''$ defined as follows: for $\forall i=1,2,...K$, and $j = 1,2,..., K$, $j\neq i$, a directed edge $(n_i -> n_j)\in E''$ exists \emph{iff} there exist distinct connected nodes $r_1, r_2, ..., r_\alpha \in H'_i$ and node $m \in H'_j$ such that $(r_1,m)\in E, (r_2,m)\in E, ..., (r_\alpha,m)\in E$.
    \item[] Step 3: Build a digraph of all the strongly connected components in $G''$, preserving the edges in $G''$ that connect the components. 
\end{enumerate}

The outcome of this method is the following: the $\pi(\alpha, \beta)$-supercore partition of $G(N, E)$ has as many $\pi(\alpha, \beta)$-supercores as the number of strongly connected components of the nodes in $G''$. A node $i\in N$ is in the $\pi$-supercore indexed by $z$ \emph{iff} $i$ finds itself in the connected component $H'_v$ of $G'$ such that $H'_v$ is in the strongly connected component in $G''$ indexed by $z$.

\begin{figure}[t]
\begin{center}
\includegraphics[height=4 in]{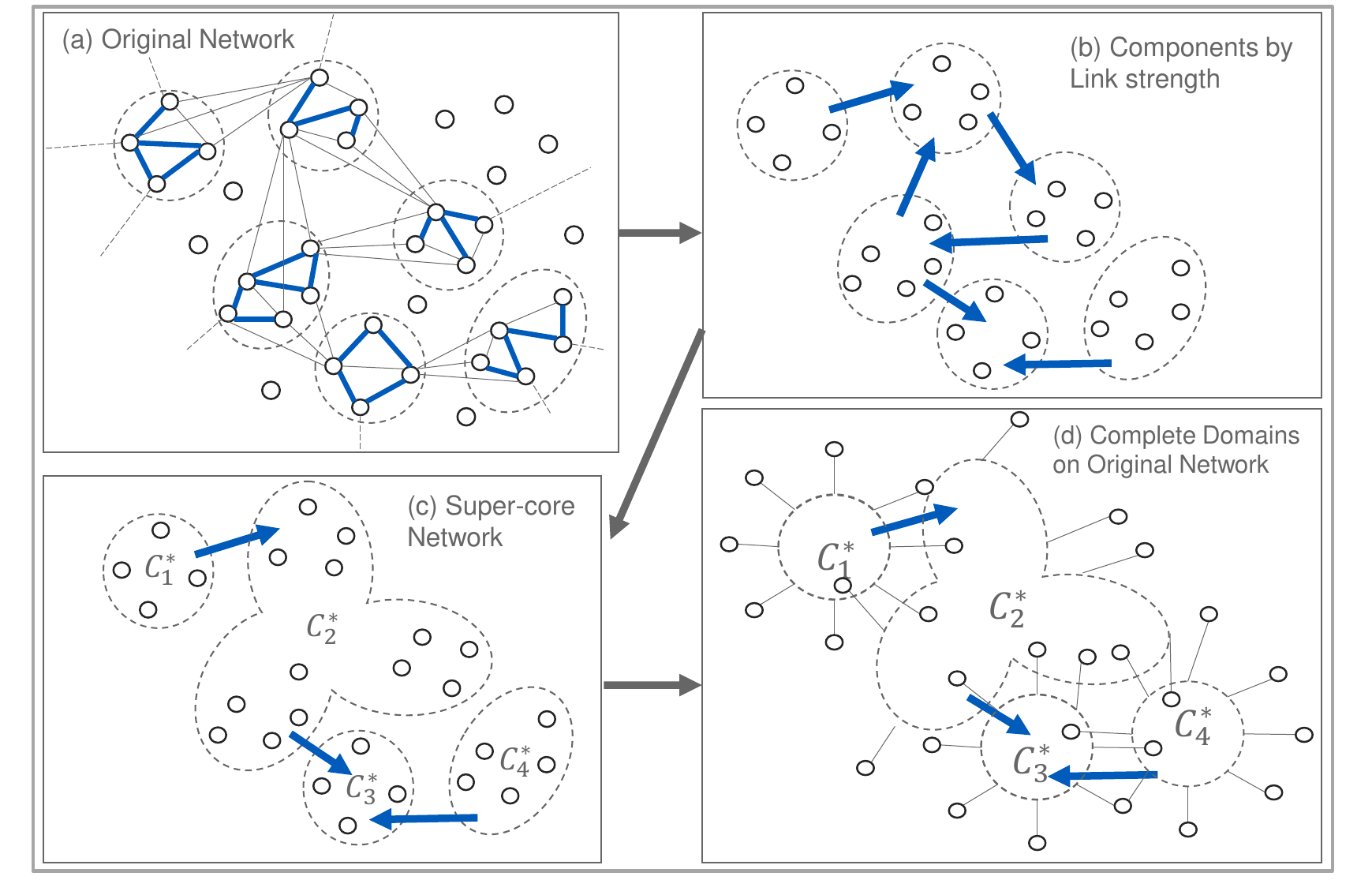}
\caption{Illustration of the method for composing $\pi(\alpha, \beta)$-compliant complete domain from graph $G$ using the directed graph of RCP-supercore. Figure (a) shows an original social graph; in it, the $\beta$-strong ties are highlighted. These strong ties form new, \emph{strongly} connected components of the graph shown using dotted circles. Figure (b) shows these components in the original network. The components have the nodes from original network $G$ (the edges are omitted in this simplistic view); special connections between the components (edges in the RCP-supercore-digraph) are represented by arrows. Figure (c) shows that all the components in (b) that are part of cycles are merged together; the resultant elements of the graph are all $\pi$-supercores. Figure (d) shows how one can compose $\pi$-compliant complete domains from the graph in (c): the directed paths indicate which $\pi$-supercores form the largest $\pi$-compliant backbones for the nodes of the original graph that they (the $\pi$-supercores) contain.}\label{Figure_4}
\end{center}
\end{figure}

We explain the steps of the method by a sequence of pictures. Figure \ref{Figure_4}(a) shows an original network, in which $\beta$-strong ties are highlighted. The strong ties define the \emph{components} shown in dotted circles; this completes Step 1. Figure \ref{Figure_4}(b) shows these components connected with directed edges as defined in Step 2 (the edges of the original network are not shown). Figure \ref{Figure_4}(c) shows that all the components that form any cycles in \ref{Figure_4}(b) are merged: all the resultant nodes are RCP-supercores, and the resultant graph is the acyclic RCP-supercore-digraph. 

Figure \ref{Figure_4}(d) illustrates how the largest RCP-compliant domains can be composed for all nodes in the original graph. Given a central node $i$, first, a RCP-compliant backbone is formed: it is comprised of all supercores reachable from $i$'s own supercore, in sequence, via the outgoing (blue) edges in the RCP-supercore-digraph. For example, in Figure \ref{Figure_4}(d), the RCP-supercore $C^*_3$ has no outgoing edges, which means that $C^*_3$ itself is the largest RCP-compliant backbone for all its nodes. The edge from $C^*_2$ to $C^*_3$ indicates that $C^*_2 \cup C^*_3$ is the largest RCP-compliant backbone for all nodes in $C^*_2$. Similarly, $C^*_1 \cup C^*_2 \cup C^*_3$ is the largest RCP-compliant backbone for all nodes in $C^*_1$. Finally, once all the backbones are found, the complete domains are obtained as the unions of the friendship circles of the nodes in these respective backbones.

Note that the presented method for composing $\pi$-supercore-digraphs, and hence, largest $\pi$-compliant domains for users in a social graph is fast. The method involves several steps, however, the computational complexity of these steps is low-polynomial. Specifically, on a sparse graph such as a typical SCN graph ($G$), Step 1 of the method finds the strengths of the edges in $G$ in time ($O(|N|^2 \log |N|)$) \citep{Itai1978}. Merging the strongly connected nodes into components that become nodes in $G''$ (as in Figure \ref{Figure_4}(b)) can be accomplished by a Breadth-First-Search (BFS) on $G$ in time $O(|N|+|E|)$. 
Step 3 requires finding all the strongly connected components in the graph $G''$, which takes linear time using.  
A BFS run on the $\pi$-supercore-digraph (as in Figure \ref{Figure_4}(c)) composes all the $\pi$-compliant backbones, and another BFS run on the original graph (as in Figure \ref{Figure_4}(d)) produces the $\pi$-compliant complete domains.

\section{Data-Driven Exploratory Analyses}
The previous sections built the theory and methods for composing domains using RCP, where RCP $\pi(\alpha,\beta)$-compliant domains of good citizens are proven disinformation-resilient, with the parameters $\alpha,\beta$ set to meet assumptions $A$1-3.

\begin{figure}[t]
\begin{center}
\includegraphics[width=17cm, height = 5.5cm]{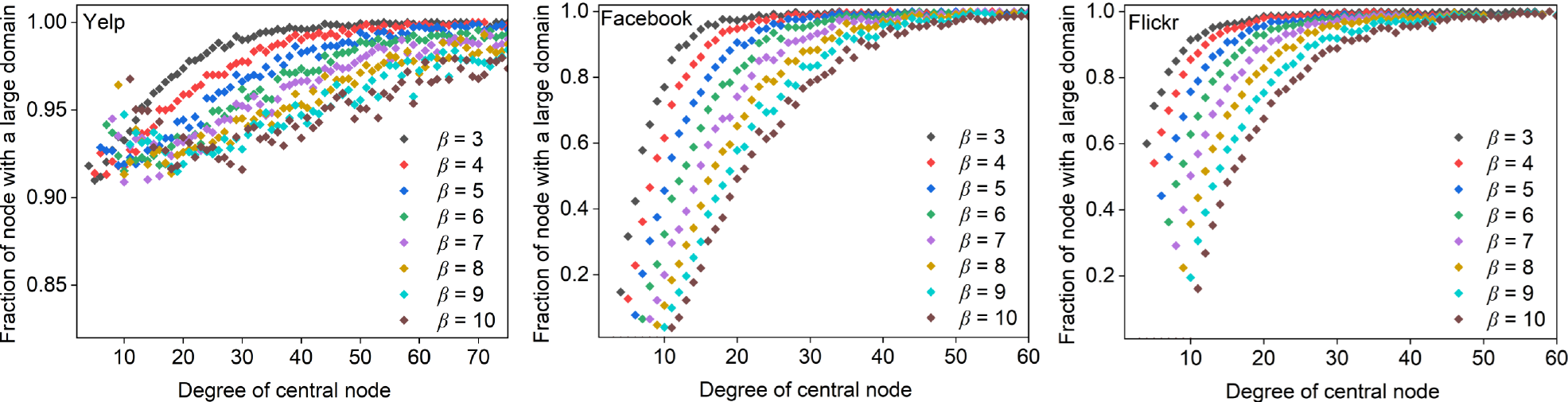}
\caption{Fraction of all nodes of the network of each dataset with fixed degree that have large $\pi(\alpha, \beta)$-compliant complete domains (of size of at least $1000$ nodes) composed for them. The RCP policy parameters are varied: $\beta = 3,4,...,10$ and $\alpha = \beta +1$.}\label{Figure_5}
\end{center}
\end{figure}

Now we apply the RCP policy with different parameters to the excerpts of real-world social networks
to explore the following questions: (1) Are RCP-compliant domains composed with real-world social networks large? (2) How does the willingness of users to provide their social connection information affect the ability of the RCP policy to produce large domains? (3) Are the RCP-compliant largest supercores, found for userbase(s) of social network(s), geospatially diverse? 

Section \ref{analysis_A} below works with social graph excerpts of Flickr, Facebook and Yelp, and describes the properties of RCP domains found for all the nodes. 
Section \ref{analysis_B} works with an excerpt of the social network of VK.com; with these data, we study the populations of the largest RCP-supercores, obtained using RCP policies with a range of values of parameter $\beta$.

\begin{table}
\begin{center}
\caption{Basic statistics of the datasets: node and link counts, average node degree, average clustering coefficient.}\label{Table_1}
\begin{tabular}{l|l|r|r|r|r}
Dataset  & Network & Nodes & Links & Avg deg & C.C. \\
\hline
Facebook & Friendship & 63731   & 817035    & 25.64      & 0.147 \\
Flickr   & Interaction  & 269970 & 33140017  & 37.12      & 0.107 \\
Yelp & Friendship        & 1032416   & 8985774    & 17.41     & 0.088\\
\hline
\end{tabular}
\end{center}
\end{table}

\subsection{Investigations into the Size of Real-World RCP Domains}\label{analysis_A}

This section uses three social network excerpts: Facebook (a social media platform), Flickr (a photo management and sharing platform), and Yelp (an e-commerce and review platform) datasets; see Table \ref{Table_1}. For Facebook and Yelp datasets, a node is a user and a link is an undirected friendship connection. The Flickr dataset is a directed network where a link represents a relationship, e.g., followship, or expresses interest of one user in seeing, liking and commenting on another user's posted photos. Since the RCP theory is presented for undirected, unweighted social graphs, we transform the  Flicker dataset to use only the symmetric connections therein.

We analyze the sizes of $\pi$-compliant complete domains composed under RCP policies. For all three datasets, the largest found domains under RCP policy $\pi=\pi(\alpha= 4, \beta=3)$ contain at least 40\% of the nodes, indicating that RCP policy is capable of composing \emph{large} domains.

Next, we observe that the size of a $\pi$-compliant complete domain depends on the availability of link information: the more links each social actor reports for themselves, the more connections they have with peers, which makes it more likely that the nodes will have larger RCP domains built for them. Conversely, for central nodes with smaller friend circles, RCP domains can be expected to be smaller. Figure \ref{Figure_5} confirms this, showing that the nodes with higher degrees indeed more often enjoy larger $\pi$-compliant domains composed for them, in comparison with the nodes with lower degrees.

\subsection{Investigations into the Geospatial Properties of Supercores Composed by RCP}\label{analysis_B}

VK.com is the largest online social portal in Eastern Europe, primarily serving the post-Soviet space \citep{semenov2018exploring}. We analyze its excerpt from 2016, which comprises the six largest cities of the country of Kazakhstan, the 3rd largest subcommunity in VK.com with 8.7M users. For each city, we study the Percentage of the Userbase in the Largest Supercore (PULS), for RCP policies with varied parameters. 

\begin{figure}[t]
\begin{center}
\includegraphics[width=17cm,]{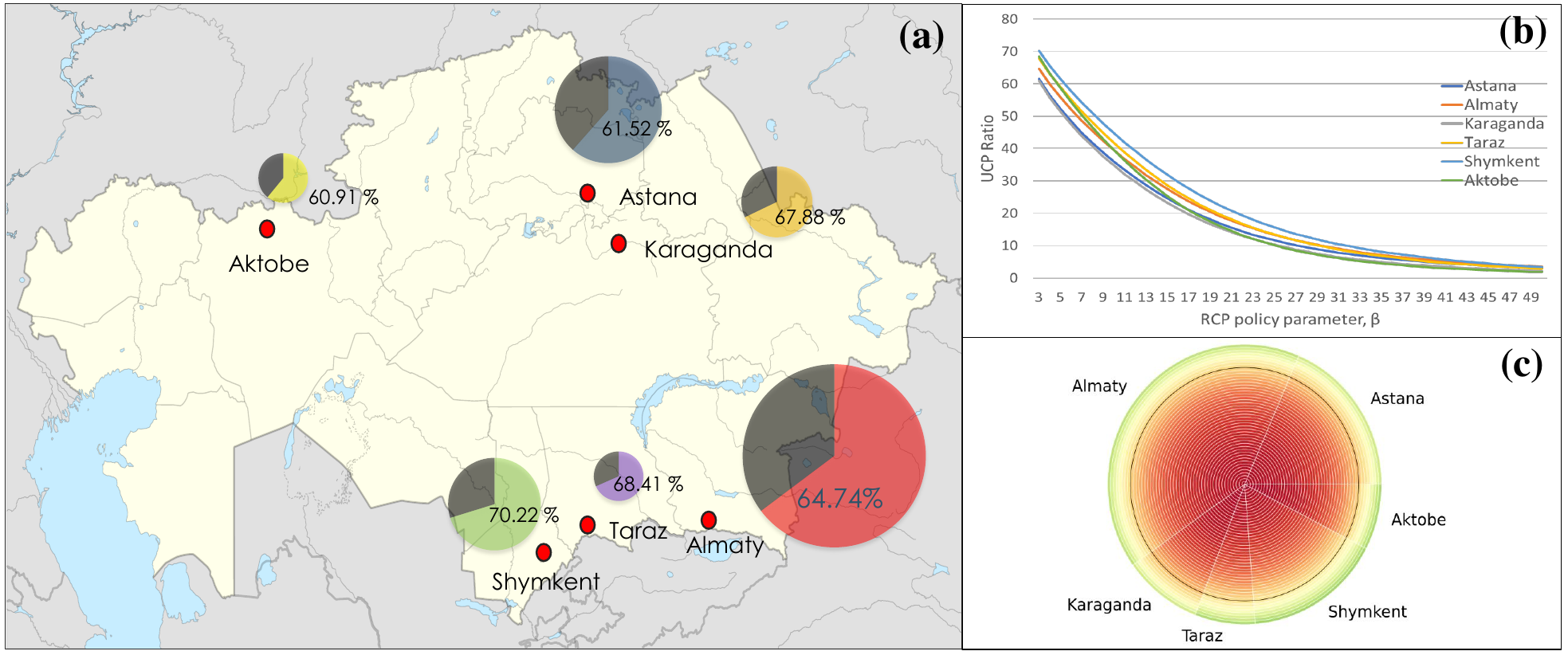}
\caption{(a) Percentage of the Userbase in the Largest Supercore (PULS), for top six cities on the Kazakhstan map, for RCP policy $\pi(4,3)$; (b) PULS values with increasing domain strengths, namely, for RCP policies with $\beta \,$=$\, 3, ... ,50$, $\alpha$=$\beta$+1; (c) Supercore heatmap: for all domain strengths, the population of users that form the largest domain is evenly spread over the cities, i.e., proportional to the city sizes.}\label{Figure_6}
\end{center}
\end{figure}

The colored segment in each pie-chart in Figure \ref{Figure_6}(a) shows what fraction of a city's VK.com users falls into the largest supercore under RCP policy $\pi(4,3)$; the pie-chart size is proportional to the city size. We find the supercore spans over the entire country and is not concentrated, e.g., in its capital.

This pattern repeats when we increase the RCP policy parameters, i.e., compose domains of higher strength. Figure \ref{Figure_6}(b) shows the PULS values for a range of $\beta$ values, with $\alpha$=$\beta$+1. 
 
Figure \ref{Figure_6}(c) gives a heatmap view on the Kazakhstan's disinformation-resilient community structure. Each segment of the disk is dedicated to a city and is crossed by 48 rings for the 48 RCP policies studied; the smaller the ring, the stricter the policy. The outer-most ring is for $\beta=3$: the resultant supercore is large but not very resilient. The inner-most ring is for $\beta=50$: this policy produces a small, highly resilient supercore. The ring segments are color-coded: green color represents high PULS values (0.702 for Shymkent for $\beta = 3$), while red represents low PULS values (0.018 for Aktobe for $\beta=50$). We find that the disinformation-resilient nucleus of the country spreads very evenly over its cities, independent of their political/territorial importance or of domain strength (level of guaranteed domain resilience), which aligns with the finding of the work of Park et al. on the strength of long-range ties \citep{Park2018}.

\section{Conclusion}
We presented a new approach to the problem of protecting SCN users from disinformation. It bypasses the need to detect bad citizens such as user accounts controlled by bots and spam farms. Instead, our approach entrusts good citizens with protecting themselves, while the SCN's information system can aggregate query response content in a personalized way. 

We presented the RCP theory and methods to compose domains, proven to be disinformation-resilient for good citizens. We showed that in practice, RCP-compliant domains are large. 

The relaxed clique percolation concept generalizes the work of \citep{Derenyi2005}: here, along with full cliques, we use relaxed cliques as percolating subgraphs. Our logic of clique-adjacency is also different. 
The main difference between the findings of \citep{Derenyi2005} and our work lies in the objectives, toward which the subgraph percolation idea is applied. The result of subgraph percolation in our case is not clusters (boundaries between global subcommunities) but node-specific subsets (domains) that are simultaneously highly flexible in structure and sufficiently cohesive to provide resiliency against disinformation attacks. 

This work will be useful for any e-commerce or review based platform that has an explicit social network structure among its users. We expect that, once platforms begin to present their users with domain-based information, e.g., showing product star-rating aggregated domain-wide (versus platform-wide), then users will see value in being more responsible about keeping their friend circles bot-free, and will report more friendships online to help the disinformation-resiliency cause. 

In extensions of this work, one could tune the RCP policy parameters for a given social graph, and perhaps, let the parameters vary for different parts of the graph. Such parameter tuning might depend on factors such as size of the network, structure and cohesiveness of the network, resiliency requirement of the platform, social link availability, etc.
Further, informational content that domains and RCP-supercores collectively contain can be compared and explained, which can aid the research on recommender systems.

\newpage
\begin{appendices}

\section{Proofs of Lemmas and Theorems}
This supplement contains proofs of the key lemmas and theorems presented in the main body of the paper. 

\subsection*{Proof of Lemma 1}
The proof is by induction. Observe first that by definition, a $\pi$-compliant backbone, $\Bar{D}^{(h)}$, composed for a good citizen $h \in H$, is a sequential expansion such that $\mathbb{X}^{\circlearrowright \textbf{P}^{\pi}}(D^{(h)}_1) = \Bar{D}^{(h)}$, specified by an ordered set of specification duplets $((R_1,Q_1), (R_2,Q_2), ...)$. To prove the statement, it is sufficient to show that, to be feasible, i.e., non-empty, each expansion in the sequence must have only good citizens in each of the candidate sets $\{Q_1, Q_2, ... \}$. 
The ``Proof by Induction" is structured as follows. We start with the smallest possible $\pi$-compliant backbone $D^{(h)}_1 = \{h\}$. The statement holds true for $D^{(h)}_1 = \{h\}$, since $h \in H$. Next, we prove that the statement holds for the \emph{first expansion} of $\mathbb{X}^{\circlearrowright \textbf{P}^{\pi}}(D^{(h)}_1)$. Then, we show that the statement holds for the \emph{$k+1$-th expansion}, given that the statement is true for $k$-th expansion.

For the \emph{first expansion}, the sentinel set $R$ will have one member only. Each node of candidate set $Q$ must have at least $\beta$ mutual friends with $h$. Since, $h \in H$, under assumption $A$2, each node in $Q$ is also a good citizen, given that $\beta \geq x$. Therefore, the statement holds for the \emph{first expansion} of $\mathbb{X}^{\circlearrowright \textbf{P}^{\pi}}(D^{(h)}_1)$.

Now, consider the statement is true for $k$-th expansion, i.e., all nodes added to $\Bar{D}^{(h)}$ upto $k$-th expansion are good citizens. Following the same argument as for \emph{first expansion}, observe that the statement holds for an expansion involving a candidate set $Q$ where each node in $Q$ has $\beta$ mutual friends with a good citizen from $R$, $\beta \geq x$. Under policy $\pi$, another feasible $k+1$-th expansion is the one where the sentinel set $R$ has at least $\alpha$ nodes that form a connected set and altogether have a common friend in candidate set $Q$. Since all nodes in $R$ are good citizens, under assumption $A_3$, the node in $Q$ is also a good citizen, given that $\alpha \geq y$. Thus, the lemma's statement holds for any expansion of $\mathbb{X}^{\circlearrowright \textbf{P}^{\pi}}(D^{(h)}_1)$ as long as $\alpha \geq y$ and $\beta \geq x$.

\subsection*{Proof of Theorem 1}
By definition of $\pi$-compliant domain, there exists a domain $\Bar{D}^{(h)}$ that is a $\pi$-compliant backbone such that $D^{(h)} = \bigcup_{m\in \Bar{D}^{(h)}} \big( m \cup F(m) \big)$. By definition of domain resiliency, it is sufficient to prove that the expected fraction of bad citizens in $\pi$-compliant complete domain $D^{(h)}$ is below the ``responsibility" threshold $r$, i.e., $\frac{\mathbb{E}(\sum_{j\in D^{(h)}} \mathbf{1_B}(j))}{|D^{(h)}|} < r$, where $\mathbf{1_B}(j) = 1$ if $j\in B$ and 0 otherwise; $0 \leq  r<< 1-r \leq 1$.

Under assumption $A$1, the expected fraction of bad citizens in a friend-circle-restricted domain (trivial domain) of any good citizen is lower than $r$. Lemma 1 guarantees that $\Bar{D}^{(h)}$ contains only good citizens under assumptions $A$1-3, given that $\alpha \geq y, \beta \geq x$. Hence, we see that $D^{(h)}$ is composed of the neighborhoods of exclusively good citizens, and calculate the expected fraction of bad citizens in $D^{(h)}$ as follows:
\begin{equation} 
\begin{split}
    \frac{1}{|D^{(h)}|} \cdot \mathbb{E}\Big(\sum_{j\in     D^{(h)}} \mathbf{1_B}(j) \mid i\in H\Big) & \\
    & = \frac{\sum_{m\in \Bar{D}^{(h)}} \Big( \mathbb{E}\Big(\mathbf{1_B}(m) \mid m\in H\Big) + \sum_{j\in F(m)} P(j\in B \mid m\in H, (i,j)\in E )\Big)} {\sum_{m\in \Bar{D}^{(h)}} \big(|F(m)|+1 \big)}\\
    & = \frac{\sum_{m\in \Bar{D}^{(h)}} \Big( 0 + r\cdot |F(m)|\Big)}    {\sum_{m\in \Bar{D}^{(h)}} \big(|F(m)|+1 \big)} = \frac{r\cdot \sum_{m\in \Bar{D}^{(h)}} \big(|F(m)|\big)}{\sum_{m\in \Bar{D}^{(h)}} \big(|F(m)|+1 \big)} < r.\\
\end{split}
\end{equation}

\subsection*{Proof of Theorem 2}
First, we prove that the intersection of any two distinct $\pi$-supercores is empty. We proceed by contradiction. Suppose there exist two $\pi$-supercores, $C^{\pi^*}_1, C^{\pi^*}_2\in N$, $C^{\pi^*}_1\neq C^{\pi^*}_2$, such that $C_{12} \equiv C^{\pi^*}_1 \cap C^{\pi^*}_2 \neq \emptyset$. Consider three nodes: $k\in C_{12}$, $i\in C^{\pi^*}_1\setminus C_{12}$, and $j\in C^{\pi^*}_2\setminus C_{12}$. Because $j\in \bar{D}^{(k)}$, then by definition of $\pi$-compliant backbone, there exists a sequential expansion $\mathbb{X}^{\circlearrowright \textbf{P}^{\pi}}(\{k\})$ specified by an ordered set of specification duplets $(\{k\},Q_1), (R_2,Q_2), (R_3,Q_3), ..., (R_M,Q_M)$ such that $j\in Q_M$. Hence, this exact sequential expansion can be used to expand $C^{\pi^*}_1$, which is a $\pi$-compliant backbone for $i$, to include node $j$ and obtain a larger $\pi$-compliant backbone for $i$.

Further, using the same logic, we see that the $\pi$-compliant backbone $C^{\pi^*}_2$ can be extended to include node $i\in C^{\pi^*}_1$ to obtain a larger $\pi$-compliant backbone for $j$. Since nodes $i$ and $j$ are selected arbitrarily (from $C^{\pi^*}_1$ and $C^{\pi^*}_2$, respectively), we conclude that for every node in $C^{\pi^*}_1$, the set $C^{\pi^*}_1 \cup C^{\pi^*}_2$ is a $\pi$-compliant backbone; and that for every node in $C^{\pi^*}_2$, the set $C^{\pi^*}_1 \cup C^{\pi^*}_2$ is also a $\pi$-compliant backbone. Therefore, set $C^{\pi^*}_1 \cup C^{\pi^*}_2$ is the $\pi$-core larger than and containing both the sets $C^{\pi^*}_1$ and $C^{\pi^*}_2$, which contradicts the definition of $C^{\pi^*}_1$ as a $\pi$-supercore and the definition of $C^{\pi^*}_2$ as a $\pi$-supercore. 

Second, we prove that every node $i \in N$ finds itself in some $\pi$-supercore. Observe that node $i \in N$ finds itself in at least one $\pi$-core, namely $\{i\}$. If it is not part of any other $\pi$-core, then $\{i\}$ is $\pi$-supercore by definition. Otherwise, i.e., if there exist other $\pi$-cores that contain $i$, then the largest of them is a $\pi$-supercore. Indeed, any two $\pi$-cores that contain $i$ cannot both be $\pi$-supercores, since they would then intersect at $\{i\}$, which is not possible as proven above. 

To sum up, every node in $N$ finds itself in one and only one $\pi$-supercore, which completes the proof.

\subsection*{Proof of Lemma 2}
The largest $\pi$-compliant backbone composed for $i\in N$, $i\in C^{\pi^*}_1$, contains all the nodes in $C^{\pi^*}_1$ by definition of $\pi$-supercore. Now define sets $R = \{r_1, r_2, ..., r_\alpha \}$ and $Q =  \{m\}$ and observe that $\mathbb{X}^{\textbf{P}^{\pi}}_{(R,Q)}(C^{\pi^*}_1)$ is a non-empty expansion since the specification duplet $(R,Q)$ satisfies expansion feasibility properties $(P^{\pi})$ as defined for RCP policy $\pi$. Hence, node $m$ is contained in the largest $\pi$-compliant backbone composed for node $i\in N$. Following the same logic used in the proof of Theorem 2, i.e., sequentially expanding the backbone $C^{\pi^*}_1 \cup m$ with $m$ as sentinel set, we conclude that the entire $\pi$-supercore $C^{\pi^*}_2$ is contained in the largest $\pi$-compliant backbone for node $i\in N$.

\subsection*{Proof of Theorem 3}
Given RCP policy $\pi(\alpha, \beta)$ and graph $G = G(N,E)$, suppose that the $\pi$-supercore-digraph $G^{\pi}$ contains a directed cycle $C^{\pi^*}_1 -> C^{\pi^*}_2 -> ... -> C^{\pi^*}_L -> C^{\pi^*}_1$, i.e., $(C^{\pi^*}_1, C^{\pi^*}_2), (C^{\pi^*}_2,C^{\pi^*}_3), ... ,(C^{\pi^*}_{L-1},C^{\pi^*}_L)$, $(C^{\pi^*}_L, C^{\pi^*}_1)\in E^{\pi}$. Then, taking any node $i\in C^{\pi^*}_1$, and applying Lemma 4, we first conclude that the largest $\pi$-compliant backbone composed for $i\in N$ contains all the nodes in $C^{\pi^*}_2$; then, applying Lemma 4 again, we conclude that the largest $\pi$-compliant backbone composed for $i\in N$ contains all the nodes in $C^{\pi^*}_3$ as well; and so on. More generally, taking any node in any of the $\pi$-supercores $C^{\pi^*}_1, C^{\pi^*}_2, ... , C^{\pi^*}_L$, we conclude that they all are in each other's largest $\pi$-compliant backbones, and hence, the union $C^{\pi^*}_1 \cup C^{\pi^*}_2  \cup ... \cup C^{\pi^*}_L$ is a $\pi$-core that is larger than, e.g., $C^{\pi^*}_1$, which contradicts the definition of $C^{\pi^*}_1$ as $\pi$-supercore.

\newpage
\section{Examples of Domain Expansion under RCP policy}

\begin{figure}[ht]
\begin{center}
\includegraphics[width=0.7\textwidth]{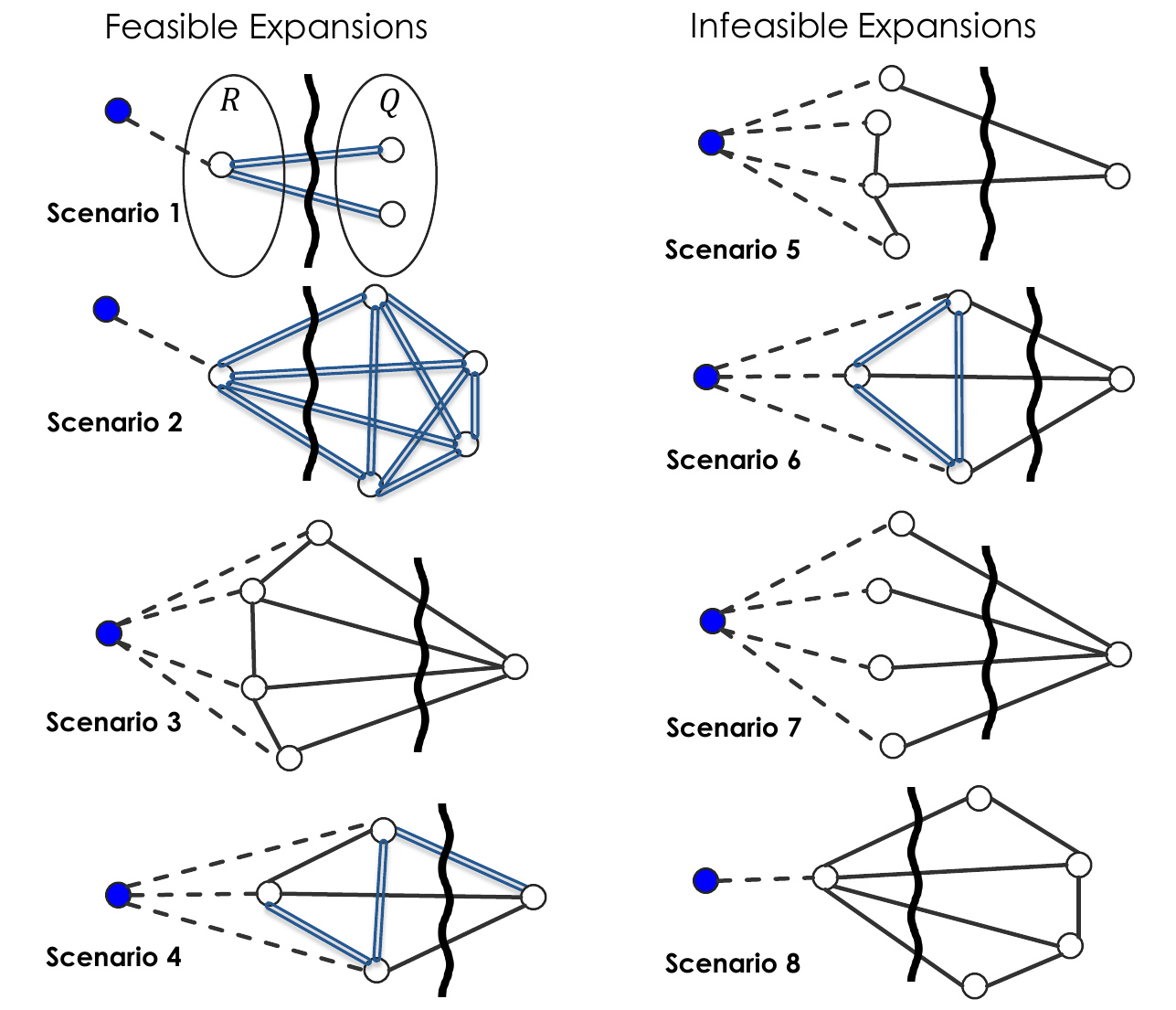}
\caption{Examples of feasible and infeasible domain expansions under the expansion feasibility properties $\textbf{P}^{\pi}=\{P^{\pi}_1, P^{\pi}_2, P^{\pi}_3\}$ defined by parametric RCP policy $\pi(\alpha,\beta)$ with parameter $\alpha=4, \beta = 3$. For each expansion depicted here -- the double blue line represents ``strong ties", i.e. atleast $\beta$ mutual friends between the connecting nodes, the vertical wavy line indicates the boundary where nodes at left side form the sentinel set $R$ and nodes at right side form the candidate set $Q$ (the sets are indicated in Scenario 1); the colored node is the central node that is  connected (represented by dashed line) with $R$. For each feasible expansion, $R\cup Q$ forms cohesive relaxed clique. Under $\textbf{P}^{\pi}=\{P^{\pi}_1, P^{\pi}_2, P^{\pi}_3\}$ defined by $\pi(\alpha=4,\beta=3)$, scenarios 1 through 4 are feasible expansions, whereas scenarios 5 through 8 are infeasible expansions.}
\end{center}
\end{figure}
\end{appendices}

\bibliography{MgtSC-ref}
\bibliographystyle{abbrvnat}
\end{document}